\documentclass{aa}  

\usepackage{graphicx}
\usepackage{txfonts}
\usepackage{orcidlink}

\usepackage{derivative}

\usepackage{subfig}

\begin{document}

    \title{Circumbinary discs as the origin of circumstellar material around interacting H-poor supernovae and fast blue optical transients}
    \titlerunning{CSM formation as CBD}

   \author{Ryotaro Chiba
          \inst{1,2}\orcidlink{0009-0003-4594-3715}\thanks{{email: ryotaro.chiba@grad.nao.ac.jp}}
          \and
          Semih Tuna\orcidlink{0000-0002-2002-6860}
          \inst{3}
          \and
          Brian D. Metzger\orcidlink{0000-0002-4670-7509}
          \inst{3,4}
          \and
          Takashi J. Moriya\orcidlink{0000-0003-1169-1954}
          \inst{2,1,5}
          }

   \institute{Astronomical Science Program, Graduate Institute for Advanced Studies, SOKENDAI, 2-21-1 Osawa, Mitaka, Tokyo 181-8588, Japan
   \and
   National Astronomical Observatory of Japan, National Institutes of Natural Sciences, 2-21-1 Osawa, Mitaka, Tokyo 181-8588, Japan
   \and
   Department of Physics and Columbia Astrophysics Laboratory, Columbia University, New York, NY 10027, USA
   \and
   Center for Computational Astrophysics, Flatiron Institute, 162 5th Avenue, New York, NY 10010, USA
   \and
   School of Physics and Astronomy, Monash University, Clayton, Victoria 3800, Australia
   }

   \date{Received XX XX, XXXX; accepted YY YY, YYYY}
 
  \abstract
   {Around 10\% of hydrogen-poor supernovae (SNe) explode inside compact ($\sim 10^{15}$ cm), massive ($\sim 0.1 \ \mathrm{M_\odot}$) circumstellar material (CSM), signalling an episode of enhanced pre-explosion mass loss whose mechanism remains unclear.
   The extreme members of this population are considered to constitute some of the fast blue optical transients (FBOTs), which exhibit rapid rise times of a few days and high peak luminosities of $\sim 10^{44} \ \mathrm{erg}$.}
   {Recent binary evolution calculations show that the expansion of helium stars during their latest evolutionary stages can trigger a rapid but stable mass-transfer episode that reaches a rate of $\sim 10^{-4} \ \mathrm{M}_\odot \ \mathrm{yr^{-1}}$ and is sustained for $\sim 10 \ \mathrm{kyr}$ before the explosion.
   It has been suggested that if this mass transfer is non-conservative and the lost mass remains gravitationally bound to the binary, it can form a dense circumbinary disc (CBD) that could explain the observed dense CSM.
   However, a detailed, quantitative analysis of this process and the resulting CBD properties such as its mass, radius, and density profile, which are required for light curve calculations and event rate estimates, has not yet been undertaken.}
   {We present a set of models that solve the viscous evolution of such a CBD subjected to time-dependent mass injection.
   This framework self-consistently includes the angular momentum exchange between the central binary and the CBD.}
   {We find that although the injected mass is initially sub-Keplerian, a lower `accretion eigenvalue' ($\chi$, corresponding to the binary torquing the CBD) prevents more mass from falling back onto the central binary.
   For our fiducial set of models with $\chi = -1$ and a viscosity parameter of $\alpha = 0.01$, the CBD immediately prior to the explosion reaches a mass of $0.07-0.20 \ \mathrm{M}_\odot$, a half-mass radius of $640 - 4000 \ \mathrm{R}_\odot$, and an aspect ratio of $\theta = H/R \sim 0.1$.
   We also show that the final CBD mass is primarily determined by $\chi$, which governs how much of the injected mass is retained within the CBD, whereas the half-mass radius is primarily determined by $\alpha$, as it controls the extent of viscous spreading.
   Finally, by calculating the evolution of interaction luminosity from a SN explosion inside the CBD, we show that the interaction between SN ejecta and the CBD can power some of the fastest-evolving interacting Type Ibc SNe that can be classified as FBOTs, such as SN 2018gep and SN 2019jc.}
   {Despite uncertainties in the model parameters, our results demonstrate that CBD formation triggered by rapid, stable mass transfer is a viable mechanism to explain the dense circumstellar environments observed around rapid, hydrogen-poor interacting SNe.
   To further investigate this CSM formation channel, a more detailed understanding of CBD--binary interaction is warranted.}


   \maketitle

\section{Introduction}

Stripped-envelope supernovae (SESNe) are explosions of stellar progenitors with little to no hydrogen-rich envelope.
The extent of stripping varies and corresponds to the spectral supernova (SN) types; from Type IIb (with moderate stripping) to Type Ib (without a hydrogen-rich envelope), and then to Type Ic (with neither H- nor He-rich envelopes; e.g. \citealt{Filippenko1997-sj, Gal-Yam2017-qs}).
While the mechanism of this envelope stripping has yet to be fully understood, there are several reasons to consider binary-specific processes, such as mass transfer \citep[e.g.][]{Podsiadlowski1992-px} or common envelope evolution \citep[e.g.][]{Paczynski1976-ps}, as an important channel.
First, the majority of massive stars reside in binary systems whose components are close enough that binary interaction will significantly affect their evolution \citep[e.g.][]{Sana2012-md}.
Furthermore, the high rate \citep[e.g.][]{Li2011-jx} and the low inferred ejecta masses \citep[e.g.][]{Drout2011-xn, Prentice2019-ef, Rodriguez2023-jl} of SESNe are inconsistent with single Wolf--Rayet progenitors being the dominant progenitor channel.

Within the SESN population, $\sim 10 \%$ show the signature of an interaction between SN ejecta and the dense circumstellar material (CSM) surrounding the progenitor \citep[e.g.][]{Ma2025-kk}.
Spectroscopically, dense CSM gives rise to characteristic narrow lines; SNe with such features are classified as Type Ibn or Icn \citep[e.g.][]{Pastorello2008-rg, Gal-Yam2022-om}.
Furthermore, the energy dissipated by the interaction enhances and sometimes dominates the light curve \citep[e.g.][]{Hosseinzadeh2017-iw}.
Alternatively, the interaction can materialise as an early bump \citep{Chiba2025-ic}.
The most extreme of these events fit the definition of fast blue optical transients (FBOTs), which are characterised by blue colours, rapid evolutionary timescales of $\sim 5$ days, and peak luminosities reaching $10^{43} \ \mathrm{erg}$ \citep[e.g.][]{Pursiainen2018-vc, Pritchard2021-zy, Leung2021-ud, Ho2023-qn, Toshikage2024-jk}.
Since CSM lies in the immediate vicinity ($\lesssim 10^{15} \ \mathrm{cm}$) of the progenitor during these explosions, its existence signifies increased mass loss shortly before the stellar explosion.
Hence, the investigation of CSM is a useful probe of pre-explosion stellar conditions.

If the progenitor is a single star, mechanisms such as gravity waves excited by core convection during the final burning stages \citep[e.g.][]{Quataert2012-ex, Fuller2017-nv, Fuller2018-ya, Wu2021-kx} or mass ejection due to pulsational pair instability \citep[e.g.][]{Woosley2007-ti, Yoshida2016-mw, Woosley2017-bo} can account for CSM formation, provided that they were triggered shortly before the explosion.
Binary interaction is another possible mechanism for CSM formation, both for hydrogen-rich SNe \citep{Ouchi2017-ri, Matsuoka2024-ho, Ercolino2024-bu} and SESNe \citep{Wu2022-ek, Ercolino2025-zx}.
The binary interaction scenario is supported by SN observations where CSM asphericity is inferred through multi-wavelength observations \citep[e.g.][]{Baer-Way2025-pq} or polarimetry \citep[e.g.][]{Mauerhan2023-iw}.
In SESNe, the binary interaction scenario is particularly compelling due to the aforementioned connection between their massive progenitors and close binaries.

In light of this, \citet{Ercolino2025-zx} conducted detailed \texttt{MESA} calculations of binary models, aiming to clarify the mass transfer history of such systems that can lead to the formation of CSM around H-poor progenitors.
They focused on the expansion of relatively low-mass He stars after the core helium burning stage in their final few thousand years \citep[e.g.][]{Paczynski1971-bi, Habets1986-nq, Yoon2012-mi, Laplace2020-il}.
By simulating the Roche lobe overflow (RLOF) and calculating the evolution of the mass transfer rate, they qualitatively estimated the properties of CSM.
They found that, if the lost mass expands in a wind-like manner with velocities as low as $\sim 10 \ \mathrm{km \ s^{-1}}$, the generated CSM extends to a radius so large that it is inconsistent with the CSM radius inferred from observation of narrow features in spectra.
However, they pointed out that, if the lost mass is gravitationally bound to the binary and forms a circumbinary disc (CBD), the resulting CSM mass and radius could be consistent with observations.

There are several reasons to believe that such a gravitationally bound CBD indeed forms in these systems.
With the high mass-loss rate ($\gtrsim 10^{-4} \ \mathrm{\mathrm{M_\odot} \ yr ^{-1}}$) that \citet{Ercolino2025-zx} found for their binary models, a large fraction of the transferred mass can leave the binary via the L2 point \citep{Lu2022-di}.
For binaries undergoing unstable mass transfer episodes, such L2 mass loss has been extensively investigated in the literature.
For example, simulations of L2 mass loss that precedes the binary coalescence find that most of the mass can indeed remain bound \citep{MacLeod2018-by, Hubova2019-wm}.
Furthermore, \citet{Scherbak2025-ru} noted in their 2D simulations of L2 mass loss triggered by a stable mass transfer episode that, while it is unclear whether the lost mass is bound, the total energy of the lost mass remains negative in their computational domain.

Circumbinary discs are widely discussed in the literature in the context of binary black holes, protoplanetary discs, planetary nebulae, and post-common envelope systems (see \citealt{Dong2024-cl} for a recent review).
In particular, a CBD whose mass is continuously fed from a  central binary has been discussed in the case of cataclysmic variables \citep{Taam2001-dv, Dubus2002-bx}.
However, a detailed investigation into CBD formation in the context of CSM formation is currently lacking.
In this work, we present a detailed analysis of such CBDs by extending the framework of \citet{Tuna2023-kh} to post-common-envelope CBD evolution, including CBD--binary interaction \citep{Siwek2023-ux, Valli2024-qi, Rocha2025-mj}.

In Sect.~\ref{sec:literature} we present the formalism used to model the viscous evolution of CBDs formed via this mechanism. 
In Sect.~\ref{sec:results} we show the results of the disc evolution for a set of models with different binary orbits, including the density structure and the spatial geometry of the CBD at the time of the explosion.
Then, we discuss the effects of key parameters, such as the viscosity parameter ($\alpha$) and the `accretion eigenvalue' ($\chi$), on the final structure of a CBD.
Finally, we predict the general behaviour of light curves arising from a SN explosion inside the CBD and the subsequent SN ejecta--CBD interaction and compare the results with observed H-poor interacting SNe, including FBOTs.
In Sect.~\ref{sec:sum} we summarise the findings of this work and discuss future outlooks.

\section{Outline of the model}
\label{sec:literature}

\subsection{Formation of the circumbinary disc}

The models in this work are based on the \texttt{MESA} binary evolution calculation of \citet{Ercolino2025-zx}.
They first evolved the single He star models to determine the detailed radius evolution of the star after the core carbon depletion.
As noted in the literature \citep[e.g.][]{Paczynski1971-bi, Habets1986-nq, Laplace2020-il}, low-mass ($\lesssim 5 \mathrm{M_\odot}$) He stars experience significant expansion during this phase, in a similar manner to main sequence stars expanding to red supergiants.
They then compared the radius evolution to a detailed binary model grid to identify which He star + main-sequence binary system could experience the Case BC RLOF triggered by said expansion.
For these models, they then ran binary evolution calculations with detailed physics sufficient to capture this RLOF process.
In their calculations, they find that the models initiates a rapid but stable mass transfer with the mass-transfer rate $\gtrsim 10^{-4} \ \mathrm{M}_\odot \ \mathrm{yr^{-1}}$ occurring $\lesssim 20 \ \mathrm{kyr}$ before explosion.
They also find that in their model, the main sequence companion spins up to the critical rotation during the Case B mass transfer, and they assume that the accretion is halted thereafter \citep{Petrovic2005-bo}.
Following their approach, we also assumed that the transferred mass does not accrete to the companion during the Case BC mass transfer.

Furthermore, we assumed that the mass lost during Case BC mass transfer remains gravitationally bound to the binary and is fully channelled into the CBD.
This is motivated by the simulations by \citet{Scherbak2025-ru}, who investigated the fate of circumbinary outflows triggered by stable mass transfer at high rates ($\gtrsim 10^{-4} \ \mathrm{M_\odot} / \mathrm{yr}$).
While they do not reach a definitive conclusion on whether the outflow ultimately remains bound due to the limited size of their computational domain, the specific total energy of the outflow is negative at the outer boundary of their simulated region (see their Fig.~10).

We adopted the mass-transfer rates calculated by \citet{Ercolino2025-zx} and treated this evolution as the mass injection term introduced at the inner edge radius of the disc in the viscous evolution equation (see Eq.~\ref{eq:mass_source}).
It should be noted that this assumption does not take into account the effect of CBD--binary interaction to the evolution of central binary.
As pointed out in \citet{Tuna2023-kh}, the stability of mass transfer can be affected by CBD--binary angular momentum exchange.
Therefore, the models with large orbital tightening are likely to be more susceptible for potential merger than the results suggest.
Furthermore, we assumed that all mass ejected from the binary remains gravitationally bound and is channelled into the CBD.
Consequently, the final CBD masses reported in this work should be treated as upper limits and order-of-magnitude estimates.
We did not consider `Case X' mass transfer during silicon burning, because the physicality of this phenomenon remains uncertain.

\subsection{Evolution of the circumbinary disc}

\subsubsection{Viscous evolution equation}

The following is based on the model devised in \cite{Tuna2023-kh}. The reader is referred to that paper for further details.

The viscous evolution of a thin, axisymmetric, Keplerian disc is derived from the conservation of mass and angular momentum within the disc.
If there is a source of mass and angular momentum, the evolution of the disc's surface density $\Sigma (r, t)$ is given by \citep{Papaloizou1995-fr}
\begin{equation}
    \frac{\partial \Sigma}{\partial t} - \frac{1}{r} \frac{\partial}{\partial r} \left( 3 \sqrt{r} \frac{\partial (\Sigma \nu \sqrt{r})}{\partial r} - \frac{2 S_\Sigma (j_\Sigma - \sqrt{G M_\mathrm{bin} r})}{\Omega} \right) = \dot{\Sigma}_\mathrm{pe} + S_\Sigma.
\end{equation}
Here, $S_\Sigma$ and $j_\Sigma$ are the injected mass per unit area and specific angular momentum, respectively.
$\dot{\Sigma}_\mathrm{pe}$ denotes the mass loss due to photoevaporation caused by irradiation from the central binary.
Note that, as shown in Sect.~\ref{sec:fiducial_set}, the effect of irradiation turned out to be nearly negligible in this model due to the small radius of CBD.
$S_\Sigma$ is given by a $\delta$-like injection from the central binary that takes the form
\begin{equation} \label{eq:mass_source}
    S_\Sigma = \frac{\dot{M}}{2 \pi R_\mathrm{in}} \delta (r - R_\mathrm{in}).
\end{equation}
We set the radius of the disc inner edge to $R_\mathrm{in} = 2 a_\mathrm{bin}$, motivated by the disc truncation due to Lindblad resonance seen in previous CBD simulations \citep{Lai2023-bt}. The
$j_\Sigma$, comparable to the Keplerian specific angular momentum at the point of mass ejection (L2), was set to $j_\Sigma = 0.7 j_\mathrm{L2}$, a value motivated by the results of recent 2D simulations \citep{Scherbak2025-ru} in the fiducial models in Sect.~\ref{sec:fiducial_set}, while different coefficients of $j_\Sigma$ are examined in Sect.~\ref{sec:parameter}.

The kinematic viscosity ($\nu$) was set using the $\alpha$-viscosity formalism:\begin{equation}
    \nu (r, \Sigma) = \alpha \frac{P}{\rho \Omega}
,\end{equation}
where $P$ is the mid-plane pressure, $\rho$ is the mid-plane density ($\rho = \Sigma / (2 H))$, and $\Omega$ is the Keplerian angular velocity.
We adopted a fiducial value of $\alpha = 0.01$ for the model presented in Sect.~\ref{sec:fiducial_set}, while exploring the dependence of the results on $\alpha$ in Sect.~\ref{sec:parameter}.
As seen later, this choice of $\alpha$ is validated since the disc is either sufficiently hot for the mid-plane to be thermally ionised ($T \gtrsim 10^3 \ \mathrm{K}$; see Fig.~\ref{fig:temperature}) or optically thin and therefore photoionised at all times, so the magnetorotational instability remains active.

We determined the disc scale height and mid-plane pressure by solving a system of equations regulating vertical hydrostatic balance, 
\begin{equation}
    P = \rho H^2 \Omega^2,
\end{equation}
where the equation of state is that for an ideal gas of particles and radiation in local thermal equilibrium, thermal balance, 
\begin{equation}
    F_\mathrm{visc} = \frac{1}{f (\tau)} (2 \sigma T^4 - F_\mathrm{irr}) + F_\mathrm{adv},
\end{equation}
with $F_\mathrm{visc}, F_\mathrm{irr}$, and $F_\mathrm{adv}$ being the heating rates per unit area due to viscosity, irradiation, and advection, respectively.
$F_\mathrm{adv}$ is given by\begin{equation}
    F_\mathrm{adv} = \frac{9}{4} \nu \Omega^2 \Sigma. 
\end{equation}
Here, $T$ is the mid-plane temperature, $\tau = \kappa \Sigma / 2$, and $\kappa$ is the Rosseland mean opacity.
Vertical radiation transfer gives the coefficient $f (\tau)$ \citep{Sirko2003-ts} as
\begin{equation}
    f(\tau) = \frac{3}{2} \tau + 2 + \frac{1}{\tau + \tau_\mathrm{min}}.
\end{equation}
We took $\tau_\mathrm{min} = 0.01$ to avoid overestimating the mid-plane in the optically thin regime.
In the simulation, $\kappa$ is evaluated with the Rosseland mean opacity law of \citet{Bell1994-yc}.
We note that this prescription assumes solar composition.
Since the CBD material is expected to be He-rich, contribution from $\mathrm{H^-}$ is replaced by $\mathrm{He^-}$, while the remaining opacity sources are expected to receive order-unity corrections due to the change in composition.
We therefore consider this prescription sufficient, as it captures the relevant $\rho$- and $T$- dependence of opacity even when for the He-rich composition.
Implementing composition-specific opacity tables is deferred to future work.

Gravitational instability within the disc is evaluated using the Toomre $Q$ parameter,
\begin{equation}
    Q = \frac{\Omega^2}{\pi G \rho}.
\end{equation}
Instability arises when $Q < Q_0$ for some $Q_0 \approx 2$.
As noted in \citet{Tuna2023-kh}, in the models, this condition is only met in regions with high disc masses, where the vertical cooling time of the disc is long compared to the dynamical time and the marginally unstable state ($Q \approx Q_0$) is achieved.
Hence, we forced the marginal stability condition $Q = Q_0 = 2$ when the instability condition was met and ignored possible fragmentation.

The initial condition of the surface density ($\Sigma$) was set in the same fashion as in \citet{Tuna2023-kh}, but here the initial total mass is negligible compared to the total mass injected during the evolution. Therefore, the choice of the initial $\Sigma$ does not affect the final results.

\subsubsection{Disc--binary angular momentum exchange}
\label{sec:am_exchange}

We assumed that the orbit stays circular and the mass ratio of the inner binary stays constant. This is a simplification since accretion from CBD into central binary is expected to drive the eccentricity towards some equilibrium value of $e \sim 0.1 - 0.5$ \citep[e.g.][]{Siwek2023-ux}.

The angular momentum exchange between the central binary and CBD can be described well by a $\delta$-like injection of angular momentum at the inner edge of CBD \citep{Rafikov2016-wq}.
Therefore, angular momentum exchange can be described as a boundary condition at the inner edge of CBD.
In this framework, the rate of net angular momentum gained by CBD $\dot{J}_\mathrm{tot}$ is given by 
\begin{align} \label{eq:am_exchange}
    \dot{J}_\mathrm{tot} &= \dot{J}_\mathrm{vis} + \dot{J}_\mathrm{adv} \\ \nonumber
    &= (3 \pi \nu \Sigma + \dot{M}_\mathrm{in}) \sqrt{G M_\mathrm{bin} R_\mathrm{in}} \\ \nonumber
    &= \chi {a_\mathrm{bin}}^2 \Omega_\mathrm{bin} \dot{M}_\mathrm{in},
\end{align}
where $\Omega_\mathrm{bin} = \sqrt{G M_\mathrm{bin} / {a_\mathrm{bin}}^3}$ and $\chi$ is a non-dimensional accretion eigenvalue that characterise how much angular momentum is transferred from the binary to the CBD \citep{Lai2023-bt}.
The limit $\chi \to - \infty$ leads to a `decretion disc' \citep{Rafikov2016-wq} with a barrier in the inner edge and is without any mass falling into the centre.
There is a significant uncertainty surrounding the value of $\chi$, and recent simulations point at its complex dependence on parameters like the binary mass ratio $q$, the binary eccentricity $e$, and the disc aspect ratio $\theta$ \citep[see e.g.][]{Munoz2020-om, Tiede2020-xk, D-Orazio2021-gc, Gagnier2023-sd, Siwek2023-ux}.
Following \citet{Tuna2023-kh}, the value of $\chi = -1$ is adopted for our fiducial models in Sect.~\ref{sec:fiducial_set}, while the models with different $\chi$ are investigated in Sect.~\ref{sec:parameter}.

Under the assumption that the binary mass ratio stays constant, binary orbital evolution due to mass and angular momentum is given by conservation of mass and angular momentum as
\begin{align}
\dot{M}_\mathrm{bin} &= - \dot{M}_\mathrm{in} - \dot{M}_\mathrm{inj}, \\
    \dot{J}_\mathrm{bin} &= \chi \dot{M}_\mathrm{in} \sqrt{G M_\mathrm{bin} a_\mathrm{bin}} - \dot{M}_\mathrm{inj} j_\Sigma, \\
    \frac{\dot{a}_\mathrm{bin}}{a_\mathrm{bin}} &= 2 \frac{\dot{J}_\mathrm{bin}}{J_\mathrm{bin}} - 3 \frac{\dot{M}_\mathrm{bin}}{M_\mathrm{bin}}, \label{eq:orbital_equation}
\end{align}
with the accretion rate into the central binary from the disc inner edge given by
\begin{equation}
    \dot{M}_\mathrm{in} = 2 \pi R_\mathrm{in} \Sigma v_\mathrm{in}(R_\mathrm{in})
,\end{equation}
where $v_\mathrm{in}(R_\mathrm{in})$ is the radial velocity of the gas with respect to the disc inner edge.
Noting that the disc inner edge was set to $R_\mathrm{in} = 2 a_\mathrm{bin}$ for a variable $a_\mathrm{bin}$, we have
\begin{equation} 
    v_\mathrm{in} (R_\mathrm{in}) = - \frac{\dot{a}_\mathrm{bin}}{a_\mathrm{bin}} R_\mathrm{in} - \frac{3}{\Sigma r^2 \Omega} {\pdv{}{r} (\nu \Sigma v^2 \Omega)}_{r = R_\mathrm{in}}.
\end{equation}

\section{Results}
\label{sec:results}

\subsection{Fiducial model set}
\label{sec:fiducial_set}

We evolved a set of disc evolution models, setting the mass injection term (Eq.~\ref{eq:mass_source}) to the RLOF mass-transfer rate evolution obtained from the binary evolution calculations of \citet{Ercolino2025-zx}.
Specifically, we adopted the values reported in their Fig.~7, which show the mass-loss evolution for models with a primary initial mass $M_i = 12.6 \ \mathrm{M_\odot}$, an initial mass ratio of $q_i = 0.5$, and initial orbital periods in the range $P_i = 5 - 63.1$ days (the calculation for the model with $P_i = 100$ days was unstable and is not included here).
We started the simulations at the onset of Case BC mass transfer and evolved them until immediately prior to the explosion.
By this stage, the primary had already lost its hydrogen-rich envelope primarily through Case B mass transfer and the its mass prior to the onset of Case BC mass transfer ranged from $2.62 - 2.97 \ \mathrm{M_\odot}$.

\begin{table*}
    \centering
    \caption{Final CBD mass ($M_\mathrm{CBD}$), final CBD half-mass radius ($R_\mathrm{CBD}$), and final binary separation ($a_f$) for different binary models.}
    \label{tab:CBD_res}
    \begin{tabular}{cccc|ccc}
        $P_i$ (d) & $a_i$ $(\mathrm{R}_\odot)$ & $\Delta M$ $(\mathrm{M}_\odot)$ & $\Delta t$ (kyr) & $M_\mathrm{CBD}$ $(\mathrm{M}_\odot)$ & $R_\mathrm{CBD}$ $(\mathrm{R}_\odot)$ & $a_f$ $(\mathrm{R}_\odot)$ \\ \hline
        5.0 & 19.5 & 0.78 & 22.5 & 0.08 & 4000 & 4.0 \\
        6.3 & 24.8 & 0.76 & 19.5 & 0.11 & 2800 & 5.5 \\
        10.0 & 38.4 & 0.71 & 10.4 & 0.17 & 1400 & 11.0 \\
        15.8 & 58.4 & 0.54 & 5.7 & 0.20 & 920 & 24.4 \\
        25.1 & 86.9 & 0.31 & 2.2 & 0.17 & 650 & 56.2 \\
        39.8 & 128.2 & 0.16 & 1.2 & 0.12 & 630 & 106.1 \\
        63.1 & 185.4 & 0.06 & 0.7 & 0.05 & 660 & 173.4 \\
    \end{tabular}
    \tablefoot{ 
    Properties of corresponding binary models are also shown: the initial binary period ($P_i$), initial separation ($a_i$), and total mass lost and duration of Case BC mass transfer ($\Delta M$ and $\Delta t$).
    The values for $P_i$, $a_i$, $\Delta M$, and $\Delta t$ are taken from Table~2 of \citet{Ercolino2025-zx}.}
\end{table*}

In Table~\ref{tab:CBD_res}, we show the final CBD total mass $M_\mathrm{CBD}$, the CBD half-mass radius $M_\mathrm{CBD}$ and the final binary separation $a_f$ along with the initial orbital period $P_i$, initial binary separation $a_i$, the total mass lost through the Case BC mass transfer $\Delta M$, and the duration of the Case BC mass transfer $\Delta t$.
We find that the final disc masses weakly depend on the initial orbital separation with values up to $0.2 \ \mathrm{M_\odot}$ for the model with $P_i = 15.8$ days.
Larger final disc radii are seen for systems with smaller initial separations and reaches $4000 \ R_\odot$ for the model with smallest initial separation.

\begin{figure}
    \centering
    \includegraphics[width=1\linewidth]{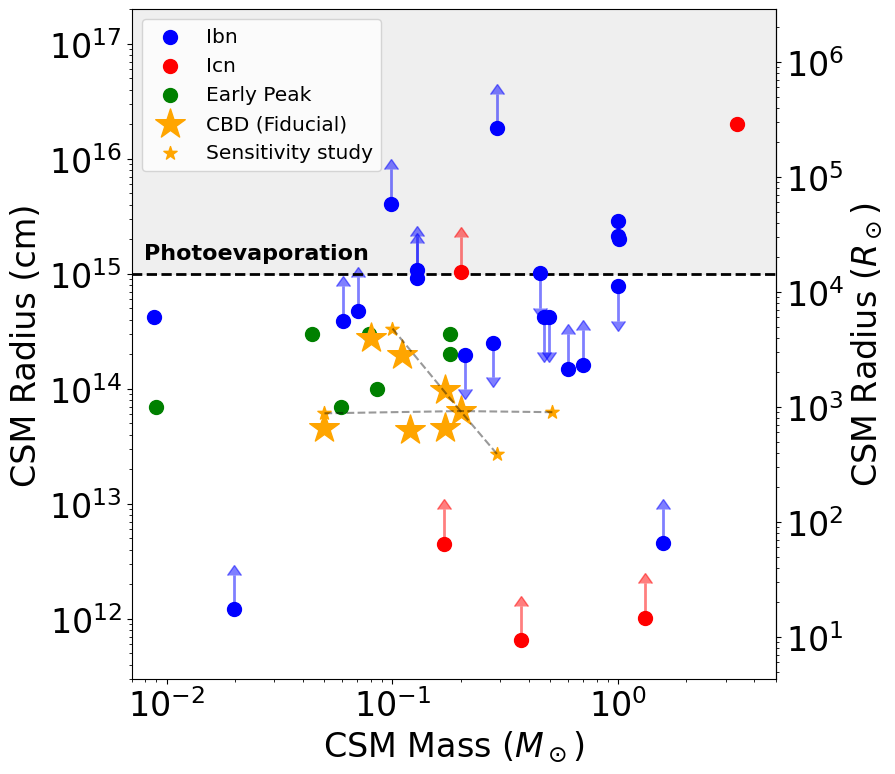}
    \caption{Comparison of final CBD masses ($M_\mathrm{CBD}$) and half-mass radii ($R_\mathrm{CBD}$) of the fiducial model suite (large stars; Table~\ref{tab:CBD_res}) and the parameter sensitivity study (small stars; Table~\ref{tab:CBD_variation}) against CSM masses and radii inferred from the observation of hydrogen-poor interacting SNe (compiled by \citealt{Ercolino2025-zx}) and hydrogen-poor SNe with early peaks \citep{Chiba2025-ic} in the literature.
    The inferred radii should be regarded as uncertain by a factor of a few.
    Note that the CSM parameter estimates quoted here generally assume the CSM is spherically symmetrical.
    We also show the range of CBD radii that is likely to be susceptible to truncation due to photoevaporation \citep{Tuna2023-kh}.
    }
    \label{fig:obs_comparison}
\end{figure}

For tighter binaries, RLOF is triggered earlier, is sustained for a longer duration, and results in greater total mass loss.
Consequently, the prolonged period of viscous spreading leads to a larger final disc radius, while the disc does not spread far enough for truncation via photoevaporation (dominant at $\gtrsim 10^{15} \ \mathrm{cm}$) to become significant.
This lack of truncation implies that the CBD radius estimate of $100 a$ adopted in \citet{Ercolino2025-zx} as an analogue to post common envelope CBD evolution is likely an overestimate.
Nonetheless, the value will provide an upper limit for the radial extent of the CBD (see also Fig.~\ref{fig:obs_comparison}).
Note that, for systems with the longest initial separations, the monotonic relation between $P_i$ and $R_\mathrm{CBD}$ breaks down due to the large inner edge radius of the CBD.

Since the specific angular momentum of the injected mass stream is sub-Keplerian, some of the material injected into the disc eventually falls back onto the central binary.
The total amount of fallback is larger for disc models with shorter initial period that evolve over longer timescales.
The competition between greater mass injection and increased mass fallback ultimately leads to the aforementioned weak dependence of the final disc mass on the initial binary separation.

Table~\ref{tab:CBD_res} also shows that models with smaller initial binary separations $a_i$ experience significant orbital tightening due to the longer duration of Case BC mass transfer and the correspondingly larger angular momentum loss via L2 mass loss.
This introduces the possibility of an eventual merger that was not seen in the original \texttt{MESA} models of \citet{Ercolino2025-zx}.

In Fig.~\ref{fig:obs_comparison}, we plot the final CBD masses and radii from Table~\ref{tab:CBD_res} alongside the CSM masses and radii inferred from observations of hydrogen-poor SNe in the literature \citep{Ercolino2025-zx, Chiba2025-ic}.
We find that, for this specific set of models, the values of the final CBD mass $M_\mathrm{CBD}$ and half-mass radius $R_\mathrm{CBD}$ are consistent with the CSM masses and radii inferred from the light curves of Type Ibc SNe exhibiting early shock-cooling peaks.
Although these events are not spectroscopically classified as interacting SNe, it makes sense that detecting an interaction signature would be difficult due to the short duration of interaction and the disc-like configuration of the CSM that can obscure the narrow lines \citep{Smith2017-lk}.

Furthermore, the small $R_\mathrm{CBD}$ are consistent with the most rapidly evolving Type Ibn/Icn SNe.
These objects can sometimes fall under the category of FBOTs due to their rapid evolution and subsequent high peak luminosities \citep{Ho2023-qn}.
In the specific case of SN 2018gep, albeit in the context of single-star mass loss models (wave-driven mass loss and pulsational pair instability), \citet{Leung2021-ud} estimated that $2 \ M_\odot$ of ejecta with an explosion energy $10^{52} \ \mathrm{erg}$ crashing into CSM of mass $\sim 0.3 \ M_\odot$ and radius of $\sim 10^3 \ R_\odot$ provides a good fit to the light curve.
We explore this comparison further in Sect.~\ref{sec:interaction_lc} by estimating the light curve arising from the ejecta--CBD interaction.

\begin{figure*}
    \sidecaption
    \begin{minipage}[b]{12cm}
        \centering
        \subfloat[\centering]
        {\includegraphics[width=5.7cm]{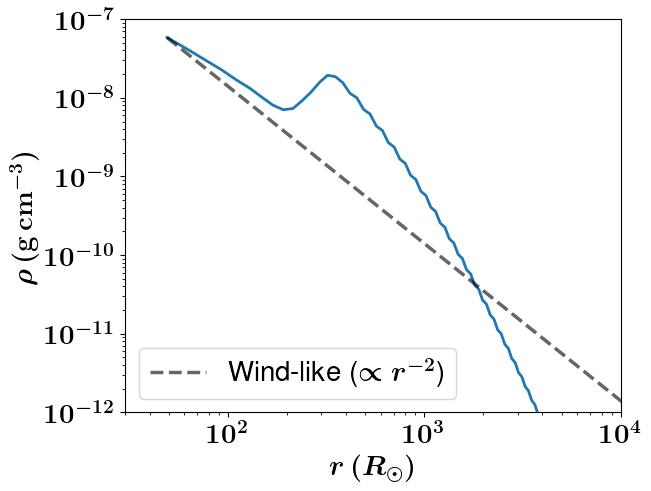}}
        \hfill
        \subfloat[\centering]
        {\includegraphics[width=5.7cm]{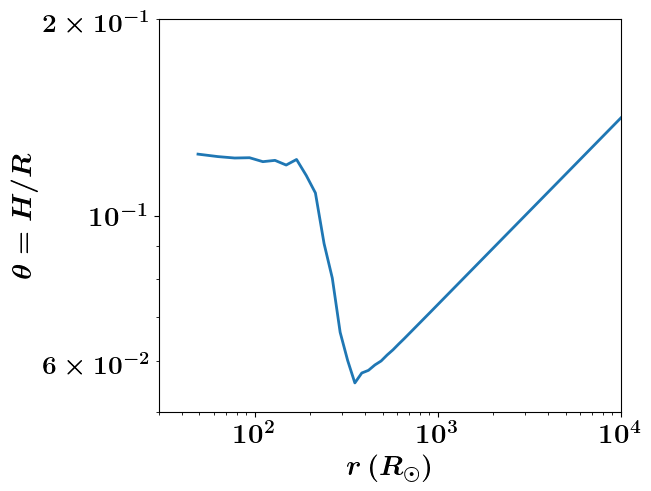}}
    \end{minipage}
    \caption{Radial profiles of the mid-plane density $\rho = \Sigma/(2H)$
    (\textit{panel a}) and aspect ratio $\theta = H/R$ (\textit{panel b}) at the end of the simulation for the
    model with $M_i = 12.6\,\mathrm{M_\odot}$, $q_i = 0.5$, and
    $P_i = 15.8$ d.}
    \label{fig:rho_theta}
\end{figure*}
\begin{figure}
    \centering
    \includegraphics[width=1\linewidth]{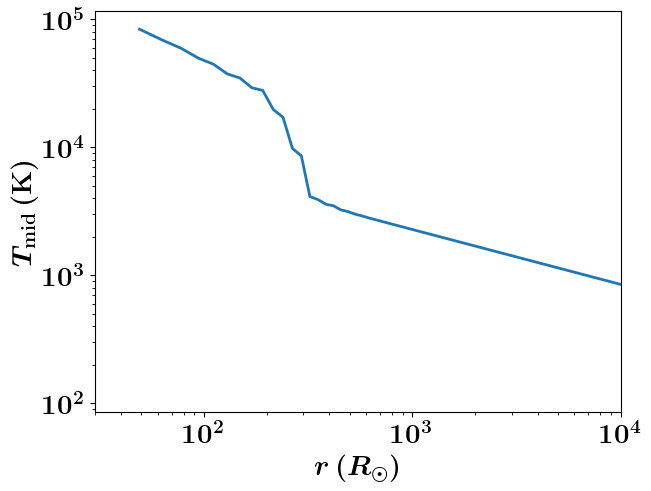}
    \caption{Mid-plane temperature for the model shown in Fig.~\ref{fig:rho_theta}}
    \label{fig:temperature}
\end{figure}

In Fig.~\ref{fig:rho_theta}, we show the final mid-plane density $\rho = \Sigma / (2 H)$ and aspect ratio $\theta = H / R$ of the CBD for the model with an initial binary orbital period of $P_i = 15.8$ d.
The overall structure is consistent with the post-common-envelope CBD solutions of \citet{Tuna2023-kh}, which show that the disc consists of three different radial zones: (1) an inner, advectively cooled, geometrically thick region, (2) an intermediate `opacity gap' transition layer where the opacity and the thickness drops, and (3) an outer, irradiation-dominated region.
In our model, this manifests as an inner zone with a nearly constant aspect ratio of $\theta \sim 0.1$, followed by a rapid decrease in $\theta$, and finally a re-flaring outer disc.
We can also see that within the half-mass radius of CBD, the disc aspect ratio stays within a narrow range of around $0.05 - 0.15$.

\subsection{Model dependence on various parameters}
\label{sec:parameter}

In this section we investigate how the properties of the CBD vary with respect to changes in several uncertain parameters.
For this analysis, we took the fiducial model with a fixed stellar masses (a primary initial mass of $M_i = 12.6 \ \mathrm{M_\odot}$ and an initial mass ratio of $q_i = 0.5$) and initial orbital period of $P_i = 15.8$ days, while varying the model parameters as follows:
\begin{itemize}
    \item The accretion eigenvalue, $\chi$ (see Eq.~\ref{eq:am_exchange} for its definition).
    Because the mass lost from the binary is initially sub-Keplerian, the value of $\chi$ determines how much angular momentum is transferred to the mass lost by the central binary and, consequently, how much material can be retained in the disc.
    The two values adopted here are $\chi = -100$, loosely corresponding to the decretion disc limit $\chi \to -\infty$ \citep{Rafikov2016-wq}, and $\chi = 0.68$, motivated by 2D viscous hydrodynamic simulations $q = 1$ binary and a $\theta = 0.1$ CBD \citep{Munoz2020-om}.

    \item The ratio ($h$) between the specific angular momentum of mass injected into the CBD ($j_\Sigma$) and the specific angular momentum at the L2 point of the binary ($j_\mathrm{L2}$).
    This also affects how much of the injected mass is retained in the disc, as it determines the extent to which the mass stream is sub-Keplerian.
    Values between $h = 0.65$ and $0.95$ are reported for binaries with different mass ratios in \citet{Scherbak2025-ru}; here, we examined the models with $h = 0.5$ and $h = 1.0$.

    \item The viscosity parameter $\alpha$.
    This parameter controls the viscous spreading of the injected mass, which was seen to be the primary factor determining the final radius of the CBD, as noted in Sect.~\ref{sec:fiducial_set}.
    Here, we adopted values of $\alpha = 0.001, 0.1$.
\end{itemize}

\begin{table}
    \centering
    \caption{Final CBD mass ($M_\mathrm{CBD}$), final CBD half-mass radius ($R_\mathrm{CBD}$), and final binary separation ($a_f$) of the models for $P_i = 15.8$ days with some specific model parameters varied.}
    \label{tab:CBD_variation}
    \begin{tabular}{c|ccc}
        Condition & $M_\mathrm{CBD}$ $(\mathrm{M}_\odot)$ & $R_\mathrm{CBD}$ $(\mathrm{R}_\odot)$ & $a_f$ $(\mathrm{R}_\odot)$ \\ \hline
        Reference & 0.20 & 920 & 24.4 \\
        $\chi = -100$ & 0.51 & 900 & 0.3 \\
        $\chi = 0.68$ & 0.05 & 880 & 48.9 \\
        $\alpha = 0.001$ & 0.29 & 390 & 26.7 \\
        $\alpha = 0.1$ & 0.10 & 4800 & 21.9 \\
        $h = 1$ & 0.24 & 940 & 20.4 \\
        $h = 0.5$ & 0.18 & 910 & 27.5 \\
    \end{tabular}
\end{table}

Table~\ref{tab:CBD_variation} and Fig.~\ref{fig:obs_comparison} present the final CBD mass ($M_\mathrm{CBD}$), half-mass radius ($R_\mathrm{CBD}$), and final binary separation ($a_f$) for the models with each parameter varied as described above.
We find that a lower value of $\chi$ significantly increases the final $M_\mathrm{CBD}$ and decreases $a_f$, without significantly altering the final $R_\mathrm{CBD}$.
The effect of $h$ is minor compared to that of $\chi$, but a higher value of $h$ results in a larger final $M_\mathrm{CBD}$ and a smaller $a_f$ is smaller, again with little impact on the final $R_\mathrm{CBD}$.
This demonstrates the consistent qualitative behaviour that the angular momentum carried by the mass stream leaving the binary determines the final $M_\mathrm{CBD}$.
In contrast, the viscosity parameter $\alpha$ is a secondary factor in determining $M_\mathrm{CBD}$ but plays a dominant role in setting $R_\mathrm{CBD}$, consistent with the interpretation that the radial extent of the disc is chiefly dictated by viscous spreading.

Another uncertain parameter is the fraction of material that remains bound after leaving the L2 point.
We assumed the entire material remains bound, for the sake of simplicity. However, \cite{Hubova2019-wm} showed that whether the L2 outflow remains bound depends on its initial condition when leaving the L2 point, while \citet{Scherbak2025-ru} did not conclusively determine whether the outflow remains bound at large radii.
\citet{Scherbak2025-ru} also noted that the radial velocity of the L2 outflow can asymptote with increasing radius at $v = 0.2 \sqrt{G M_\mathrm{bin} / a_\mathrm{bin}}$, comparable to the binary orbital velocity.
For our model with $P_i = 15.8 \ \mathrm{d}$, initial binary mass $9.27 \ M_\odot$ and separation $58.4 \ R_\odot$ gives $v = 34.8 \ \mathrm{km/s}$.
In the corresponding binary model, Case BC RLOF begins $\sim 5 \ \mathrm{kyr}$ before the explosion, giving the estimate an outer radius of unbound material $\sim 5.5 \times 10^{17} \ \mathrm{cm}$.
The unbound outflow could therefore form extended CSM of radius $\sim 10^{17} - 10^{18} \ \mathrm{cm}$, similar to what is expected in constant-velocity expansion at $10 \ \mathrm{km/s}$ of unbound material, considered in \citet{Ercolino2025-zx}.
For the RLOF rate $\dot{M} \sim 10^{-4} \ M_\odot / \mathrm{yr}$ and $v = 34.8 \ \mathrm{km \ s^{-1}}$, CSM density ($\propto \dot{M} / v$) is comparable to those estimated for SNe with radio emission detected by Very Large Array Sky Survey \citep[$\dot{M} = 10^{-4} - 10^{-1} \ M_\odot \ \mathrm{yr^{-1}}$ for $v = 1000 \ \mathrm{km \ s^{-1}}$, ][]{Stroh2021-tp}.
Therefore, interaction with this material could produce late-time radio emission and provide an observational signature of the binary scenario.

\subsection{Interaction light curve}
\label{sec:interaction_lc}

When a SN explodes inside dense CSM, the collision between SN ejecta and CSM dissipates the kinetic energy of the ejecta, converting some fraction of it into radiation.
In this section we consider the interaction between SN ejecta and a CBD, adopting the density structure and the geometry derived from the models presented in Sect.~\ref{sec:fiducial_set}, and estimate the evolution of the interaction-powered luminosity.

We adopted the standard formalism of \citet{Moriya2013-lj}, which assumes that the shocked region rapidly cools and collapses into a geometrically thin shell, i.e. a cold dense shell.
In this framework, the dynamics of the system are governed by mass and momentum conservation of the cold dense shell and the radiated luminosity is taken to be a certain fraction of the kinetic energy dissipated at the shocks.
To account for the geometry of the CBD, we extended the standard spherical formalism by solving the thin-shell equations independently along each polar angle and integrating the resulting emission over the solid angle. The detailed formulation is presented in Appendix~\ref{sec:model_equation}.

\begin{figure}
    \centering
    \includegraphics[width=1\linewidth]{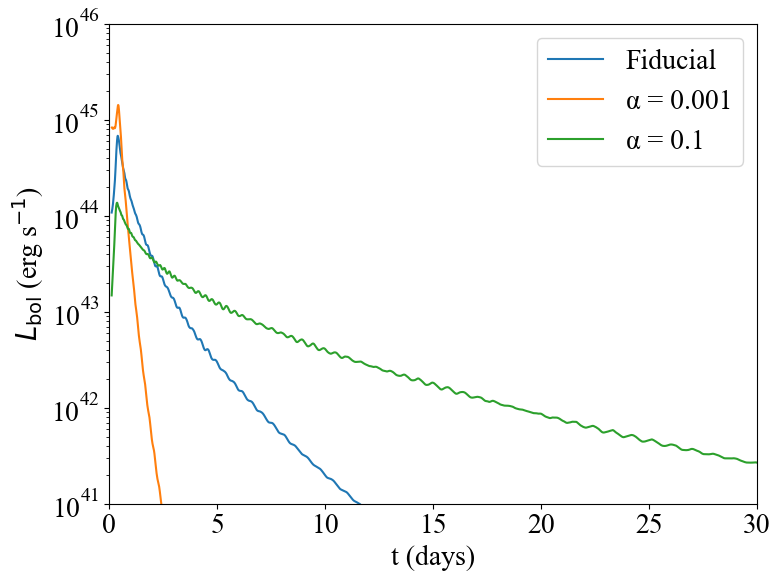}
    \caption{Evolution of the interaction luminosity.
    We consider SN ejecta with a mass of $1.42 \ M_\odot$ and an explosion energy of $10^{51} \ \mathrm{erg}$, assuming a double power-law density structure (see Appendix~\ref{sec:model_equation} for details).
    Density structures and the geometry of the CBD are taken from the models presented in Sect.~\ref{sec:results} with an initial binary period $P_i = 15.8$ d and different viscosity parameters, $\alpha$.}
    \label{fig:lc_set}
\end{figure}

Figure~\ref{fig:lc_set} shows the evolution of the bolometric interaction luminosity for three of our models.
Each model corresponds to the binary with initial period $P_i = 15.8$ d, but with different viscosity parameter $\alpha$, which, as shown previously, most strongly affects the disc radius and thus its density structure.
Since the initial mass of the He star in the fiducial model was $2.82 \ M_\odot$, we set the ejecta mass to $1.42 \ M_\odot$ (i.e. the remnant neutron star mass of $1.4 \ M_\odot$).
The explosion energy was set to $10^{51} \ \mathrm{erg}$, a typical value for Type Ib SNe \citep{Drout2011-xn}.
We find that the luminosity in these models peaks at around a few days and declines thereafter.
This behaviour is consistent with the light curve calculations of \citet{Ercolino2025-zx} for the interaction between SN ejecta and a CBD-like aspherical CSM with a density profile $\propto r^{-2}$ (our models exhibit an even steeper density decline).

This evolution is significantly more rapid than that of most Type Ibn light curves, which typically peak at $\sim 10$ days after the explosion \citep{Hosseinzadeh2017-iw}.
However, the short timescale and peak luminosities reaching $\gtrsim 10^{44} \ \mathrm{erg}$ resemble rapidly evolving interacting SNe that are sometimes designated as FBOTs \citep{Ho2023-qn}.
Indeed, previous works show that shock breakout in dense spherical CSM with radii comparable to those of our CBD models can reproduce FBOT-like light curves, for example KSN 2015K \citep{Rest2018-ah} and SN 2018gep \citep{Leung2021-ud}.

Motivated by this, we estimated the rise time following shock breakout using analytical framework \citep[e.g.][]{Chevalier2011-xw, Khatami2024-ex}.
From the mid-plane density profile in Fig.~\ref{fig:rho_theta} and the radial evolution of shock in the mid-plane obtained in the current framework, we could estimate the breakout time $t_\mathrm{br}$ as the time when $c / v_\mathrm{sh} = \tau_\mathrm{CSM}$, where $v_\mathrm{sh}$ is the velocity of shocked shell and $\tau_\mathrm{CSM} = \int_{r_\mathrm{sh}} \kappa \rho (r) \ \mathrm{d} r$ is the optical depth of unshocked CSM lying ahead of the shock.
With the fiducial value of $\kappa = 0.2 \ \mathrm{cm^2 / g}$, we obtain $t_\mathrm{br} = 5.3 \ \mathrm{d}$.
At this time, $r_{\mathrm{sh}} = 1.6 \times10^{14} \ \mathrm{cm}$, $v_{\mathrm{sh}} = 2.7 \times 10^{3}\ \mathrm{km / s}$, and $\rho_{\mathrm{CSM}} = 1.48 \times 10^{-11}\ \mathrm{g / cm}$.
The rise time of the light curve is comparable to the dynamical timescale $\Delta R / v_\mathrm{sh}$, where $\Delta R$ is the thickness of the region from which the photons breaks out.
Since the optical depth across this region, $\kappa \rho_\mathrm{CSM} \Delta R$, should be comparable to $c / v_\mathrm{sh}$, the rise time is then estimated as $\sim c / (\kappa \rho_\mathrm{CSM} {v_\mathrm{sh}}^2) \approx 1.6$ d.
Therefore, if seen from an edge-on direction, the object can have a rise time of the order of a few days.

We caution, however, that the CSM masses inferred for the aforementioned spherical models of events such as KSN 2015K and SN 2018gep cannot be compared directly to the total CBD mass in our aspherical models; an edge-on observer samples only a limited solid angle, while the large total optical depth implies that part of the dissipated energy may be redistributed into a broader, lower-luminosity component.
At shock breakout, the optical depth of the unshocked CSM in the radial direction $\tau_\mathrm{CSM}$ can be calculated as $\int_{r_\mathrm{sh}} \kappa \rho (r) \ \mathrm{d} r= 111$, whereas the optical depth in the direction perpendicular to the disc is $\kappa \rho_\mathrm{CSM} H = 43$.
The two CSM optical depths are therefore comparable, allowing some radiation to diffuse perpendicular to the disc and subsequently through the ejecta, emerging on a much longer timescale of several tens of days at lower luminosity and potentially appearing as a later excess in the light curve.

If instead the object were seen from a face-on viewing angle, radiation from interaction would diffuse through the SN ejecta. This could instead appear as an early luminosity excess, which has been reported in SESNe \citep[e.g.][]{Rodriguez2023-jl}.
Quantitative predictions of this viewing-angle dependence require multidimensional radiation-transport calculations and is deferred to future works.

SN 2018gep exhibited precursor emission for several weeks before explosion; this was previously interpreted as evidence of an eruptive mass-loss episode on a similar timescale \citep{Ho2019-be}.
The CBD in our model forms through mass transfer sustained over a much longer timescale and therefore the radiative emission from CBD cannot account for this precursor.
One possible explanation is a late episode of enhanced accretion onto the companion similar to the scenario of \citet{Tsuna2024-rt}.
Note that the CBD formation scenario considered here applies whether the companion is an MS star or a compact object.

\section{Summary and conclusions}
\label{sec:sum}

We have investigated late-stage, stable mass transfer as a mechanism for forming dense CSM around SESN progenitors. 
Motivated by recent binary evolution models, we computed the viscous evolution of the CBD formed from mass lost through the L2 point. 
Our results demonstrate that this mass-transfer channel can successfully produce a CBD with masses ($\sim 0.1 \ M_\odot$) and radii ($\sim 10^3 \ R_\odot$) consistent with the dense CSM inferred from observations of Type Ibc SNe with early peaks, as well as the most rapid population of Type Ibn/Icn SNe that constitute some of the FBOTs.
Furthermore, a potentially unbound component of the same outflow can also produce late-time radio emission similar to the known radio SNe.

One uncertainty in our models is the eventual orbital stability of the central binary.
A consequence of CBD formation is the extraction of angular momentum from the central binary, which can lead to significant orbital migration (as noted in Table~\ref{tab:CBD_res}).
As the orbit tightens, the mass-transfer rate will increase.
For the fiducial model, we estimate an increase by a factor of a few (see Appendix B for the derivation).
Therefore, for models with initially tight binaries, a merger prior to core collapse is more likely than our models suggest.
To resolve this uncertainty, a consistent treatment of binary evolution and CBD formation is required, presumably incorporating detailed treatments like the radial change in the donor due to envelope instability and the effects of eccentricity \citep{Parkosidis2026-yk, Parkosidis2026-wj}.

Furthermore, our framework relies on several parameterised treatments.
The largest uncertainties in our model stem from the viscosity parameter $\alpha$ and the accretion eigenvalue $\chi$.
While our parameter study captures the qualitative behaviour of the CBD evolution, determining the precise consequences of CBD formation will require inputs from 3D hydrodynamic simulations.

\begin{acknowledgements}
    Authors thank discussions with Andris Doroszmai and Utkarsh Jain for valuable comments regarding this work.
    This work is supported by the Grants-in-Aid for Scientific Research of the Japan Society for the Promotion of Science (JP24K00682, JP24H01824, JP21H04997, JP24H00002, JP24H00027, JP24K00668) and by the Australian Research Council (ARC) through the ARC's Discovery Projects funding scheme (project DP240101786).
    ST and BDM are supported in part by NASA (grants 80NSSC22K0807, 80NSSC24K0408) and the Simons Foundation (grant 727700).
    The Flatiron Institute is supported by the Simons Foundation. 
\end{acknowledgements}

\bibliographystyle{aa}
\bibliography{paperpile}

\begin{appendix}
\section{Formalism for computing the ejecta–CBD interaction luminosity}
\label{sec:model_equation}

We assumed that the SN ejecta expands homologously and that its density structure can be described with a double power law \citep{Matzner1999-lt}:
\begin{equation} \label{eq:ejecta_density}
    \rho_\mathrm{ej} (v, t) \propto
    \begin{cases}
        t^{-3} (v / v_\mathrm{tr})^{- \delta} & (v > v_\mathrm{tr}) \\
        t^{-3} (v / v_\mathrm{tr})^{- n} & (v < v_\mathrm{tr})
    \end{cases}
\end{equation}
with
\begin{equation}
    v_\mathrm{tr} = \sqrt{\frac{2 (5 - \delta) (n - 5) E_\mathrm{ej}}{(3 - \delta)(n - 3) M_\mathrm{ej}}}.
\end{equation}
Here $E_\mathrm{ej}$ and $M_\mathrm{ej}$ are the total kinetic energy and total mass of the SN ejecta, respectively.
Numerical simulations suggest that the typical values of power indices are $\delta = 0 - 1$ and $n = 7 - 12$; we adopted $\delta = 0$, $n = 10$.

This SN ejecta collides with CBD that is radially stationary and has the axisymmetric density distribution described by
\begin{equation}
    \rho_\mathrm{CBD} (r, \theta) \propto
    \begin{cases}
        \rho_\mathrm{mid} (r) & (|\theta - \pi/2| < \theta_\mathrm{d}(r)) \\
        0 & (|\theta - \pi/2| > \theta_\mathrm{d}(r)).
    \end{cases}
\end{equation}
The $\rho_\mathrm{mid} (r)$ and $\theta_\mathrm{d}(r)$ are the radial distribution of the mid-plane density and the opening angle corresponding to the disc aspect ratio, respectively.

We adopted the thin-shell approximation, treating the shocked region between the forward and reverse shocks as a geometrically thin shell characterised by its radius $r_\mathrm{sh} (t, \theta)$ and mass per unit solid angle $m_\mathrm{sh} (t, \theta)$.
We determined the evolution of the shell  independently at each polar angle ($\theta$) by solving the conservation of mass and momentum.
Using velocity of the shell $v_\mathrm{sh} \coloneqq  \pdv{r_\mathrm{sh}}{t}$, they can be described as
\begin{equation}
    \odv{m_\mathrm{sh}}{t} = {r_\mathrm{sh}}^2 [\rho_\mathrm{ej} (v_\mathrm{ej} - v_\mathrm{sh}) + \rho_\mathrm{CBD} v_\mathrm{sh}],
\end{equation}
and
\begin{equation}
    m_\mathrm{sh} \odv{v_\mathrm{sh}}{t} = {r_\mathrm{sh}}^2 [\rho_\mathrm{ej} (v_\mathrm{ej} - v_\mathrm{sh})^2 - \rho_\mathrm{CBD} {v_\mathrm{sh}}^2].
\end{equation}

The instantaneous interaction luminosity per unit solid angle is calculated from the kinetic energy dissipated at the forward and reverse shocks,
\begin{equation}
    \odv{L_\mathrm{int}}{\Omega} = \frac{1}{2} \varepsilon {r_\mathrm{sh}}^2 [\rho_\mathrm{ej} (v_\mathrm{sh} - v_\mathrm{ej})^3 + \rho_\mathrm{CBD} {v_\mathrm{sh}}^3] \label{eq:lum_int_general}
,\end{equation}
where $\varepsilon$ is the conversion efficiency from kinetic to radiation energy.
We used $\varepsilon = 0.3$ here, motivated by observation \citep[e.g.][]{Fransson2014-tv}.
The total bolometric luminosity is then obtained by integrating over all solid angles:
\begin{equation}
    L_\mathrm{int} (t) = 2 \pi \int_0^\pi \odv{L_\mathrm{int}}{\Omega} (t, \theta) \sin \theta \ \mathrm{d} {\theta}.
\end{equation}

\section{Estimate of the uncertainty in the mass transfer rate due to orbital evolution}

We estimated how CBD-driven orbital contraction changes the RLOF rate.
The estimate treats the orbital tightening from the CBD calculation (Table~\ref{tab:CBD_variation}) as an additional driver of mass transfer.

The evolution of donor radius ($R_\mathrm{d}$) can be separated into its response to mass loss and intrinsic stellar evolution as
\begin{equation}
    \frac{\dot R_\mathrm{d}}{R_\mathrm{d}} = \zeta_\mathrm{s} \frac{\dot M_\mathrm{d}}{M_\mathrm{d}} + \left(\frac{\partial\ln R_\mathrm{d}}{\partial t}\right)_\mathrm{nuc}
,\end{equation}
where $\zeta_\mathrm{s}$ is the adiabatic mass--radius exponent \citep{King1996-be}. 
For the evolution of the Roche lobe radius, we separated the response already present in the original binary model and the contribution from additional orbital evolution due to binary--CBD interaction.
Since $R_\mathrm{L}=f(q)a_\mathrm{bin}$ in the \citet{Eggleton1983-nt} prescription and the binary mass ratio ($q$) is assumed to be constant, the additional CBD contribution satisfies $(\dot R_\mathrm{L}/R_\mathrm{L})_\mathrm{CBD}=(\dot a_\mathrm{bin}/a_\mathrm{bin})_\mathrm{CBD}$.
Therefore, 
\begin{equation}
    \frac{\dot R_\mathrm{L}}{R_\mathrm{L}} = \zeta_\mathrm{R} \frac{\dot M_\mathrm{d}}{M_\mathrm{d}} + \left( \frac{\dot a_\mathrm{bin}}{a_\mathrm{bin}} \right)_\mathrm{CBD}.
    \label{eq:app_RL_evol}
\end{equation}
Here, $\zeta_\mathrm{R}$ is the Roche-lobe mass--radius exponent of the original binary model.

Since $R_\mathrm{d} = R_\mathrm{L}$ for steady mass transfer and $\dot M_\mathrm{RLOF} = - \dot M_\mathrm{d}$, we have
\begin{equation}
    (\zeta_\mathrm{s} - \zeta_\mathrm{R}) \frac{\dot M_\mathrm{RLOF}}{M_\mathrm{d}} = \left(\frac{\partial\ln R_\mathrm{d}}{\partial t}\right)_\mathrm{nuc} - \left(\frac{\dot a_\mathrm{bin}}{a_\mathrm{bin}}\right)_\mathrm{CBD}.
    \label{eq:app_contact_cbd}
\end{equation}
The RLOF rate in the original model $\dot M_\mathrm{RLOF,0}$ without the contribution from additional orbital evolution would have been
\begin{equation}
    (\zeta_\mathrm{s} - \zeta_\mathrm{R}) \frac{\dot M_\mathrm{RLOF, 0}}{M_\mathrm{d}} = \left(\frac{\partial\ln R_\mathrm{d}}{\partial t}\right)_\mathrm{nuc}
    \label{eq:app_contact_original}
\end{equation}
and therefore
\begin{equation}
    \frac{\dot M_\mathrm{RLOF}}{\dot M_\mathrm{RLOF,0}} = 1 - \frac{(\dot a_\mathrm{bin}/a_\mathrm{bin})_\mathrm{CBD}}{({\partial\ln R_\mathrm{d}} / {\partial t})_\mathrm{nuc}}.
    \label{eq:app_mdot_ratio}
\end{equation}

We estimated the fractional nuclear expansion rate $({\partial\ln R_\mathrm{d}} / {\partial t})_\mathrm{nuc}$ from the binary models of \citet{Ercolino2025-zx}.
Their Table~2 gives two binary models with the same initial primary mass of $12.6 \ M_\odot$, initial periods of $15.9$ and $25.1\,\mathrm{d}$, and Roche lobe radii of $18.1$ and $27.1 \ R_\odot$; their Case BC durations are $5.70$ and $2.24 \ \mathrm{kyr}$.
Since donors are detached before the onset of Case BC mass transfer, we treated the companions as following nearly identical intrinsic stellar radius evolution.
Thus, we could approximate the fractional nuclear expansion rate as
\begin{equation}
    \left(\frac{\partial\ln R_\mathrm{d}}{\partial t}\right)_\mathrm{nuc} \sim \frac{\ln(27.1/18.1)}{(5.70-2.24)\times10^3 \ \mathrm{yr}} = 1.17\times10^{-4} \ \mathrm{yr}^{-1}.
    \label{eq:log_expansion_rate}
\end{equation}
We also estimated the logarithmic orbital migration rate using values from Table~\ref{tab:CBD_res} for the fiducial $P_\mathrm{i}=15.8 \ \mathrm{d}$ model as
\begin{equation}
    \frac{\dot a_\mathrm{bin}}{a_\mathrm{bin}} \sim \frac{\ln(24.4/58.4)}{5.70\times10^3\,\mathrm{yr}} = -1.53\times10^{-4}\,\mathrm{yr}^{-1}.
    \label{eq:app_adot_num}
\end{equation}
Therefore, we estimate
\begin{equation}
    \frac{\dot M_\mathrm{RLOF}}{\dot M_\mathrm{RLOF,0}} \sim 1+\frac{1.53\times10^{-4}}{1.17\times10^{-4}} = 2.31,
    \label{eq:app_mdot_result}
\end{equation}
i.e. a factor-of-a-few increase in the mass transfer rate.

\end{appendix}

\end{document}